\documentclass[aps,preprint,amsmath,amssymb]{revtex4-1}
\usepackage{graphicx}

\newcommand{\bd}{\begin{document}}
\newcommand{\ed}{\end{document}}
\newcommand{\bc}{\begin{center}}
\newcommand{\ec}{\end{center}}
\newcommand{\be}{\begin{eqnarray}}
\newcommand{\ee}{\end{eqnarray}}
\newcommand{\ba}{\begin{array}}
\newcommand{\ea}{\ed{array}}
\newcommand{\strich}[1]{#1  \! \! \slash}
\newcommand{\eqn}{\global\def\theequation}
\newcommand{\sw}{sin^2 \theta_W}
\newcommand{\fbd}{f_B}
\renewcommand{\thefootnote}{\alph{footnote}}
\newcommand{\se}{\section}
\newcommand{\sse}{\subsection}
\newcommand{\bi}{\bibitem}
\def\figcap{\section*{Figure Captions\markboth
     {FIGURECAPTIONS}{FIGURECAPTIONS}}\list
     {Figure \arabic{enumi}:\hfill}{\settowidth\labelwidth{Figure 999:}
     \leftmargin\labelwidth
     \advance\leftmargin\labelsep\usecounter{enumi}}}
\let\endfigcap\endlist \relax
\def\reflist{\section*{References\markboth
     {REFLIST}{REFLIST}}\list
     {[\arabic{enumi}]\hfill}{\settowidth\labelwidth{[999]}
     \leftmargin\labelwidth
     \advance\leftmargin\labelsep\usecounter{enumi}}}
\let\endreflist\endlist \relax

\begin{document}
\title
{\Large {\bf Do leptoquarks manifest themselves in ultra-high energy neutrino interactions?}}

\author{ I. Alikhanov}
\email[]{E-mail: ialspbu@gmail.com}

\affiliation{Institute for Nuclear Research of the Russian Academy of Sciences,
60-th October Anniversary pr. 7a, Moscow 117312, Russia
 }


\begin{abstract}
It is assumed that neutrino--nucleon scattering at ultra-high energies effectively proceeds through excitations of leptoquarks in neutrino--quark subprocesses. This approach reproduces the behavior of the energy dependence of the ultra-high energy neutrino--nucleon scattering cross sections and allows to estimate masses as well as the decay widths of the involved leptoquarks. For instance, this leads to the leptoquark mass $1353\pm230$ GeV in a way independent on the leptoquark quantum numbers. The discovery potential of the LHC for the leptoquarks is evaluated.

\end{abstract}

\maketitle %

\section{Introduction}
\label{intro}
There are reactions whose properties can be described in terms of effective interactions. Such interactions allow to reduce the number of degrees of freedom in comparison with direct calculations and thus considerably simplify the problem of finding the cross sections. One of the clearest examples in particle physics is the vector meson dominance model. The cross sections of high-energy photon interactions with nucleons and nuclei were observed to resemble that of purely hadronic reactions but smaller by about a factor of $\alpha=1/137$. Relying on this observation  it was assumed that the photon in $\gamma N$ reactions transforms into a vector meson $\rho$, $\omega$, $\phi$ and then interact with the target, so that the $\gamma N$ reactions effectively proceed as hadron--hadron ones. Another example is provided by the Regge theory which reproduces a number of properties of hadron--hadron scattering by assuming that the hadrons interact with each other through the exchange of a family of particles called reggeons.

The present paper shows that the energy dependence of the ultra-high energy (UHE) neutrino--nucleon scattering cross sections can be reproduced if one assumes that the UHE neutrinos effectively interact with the nucleons through excitations of leptoquarks  in neutrino--quark subprocesses. The standard model, within which the UHE neutrino--nucleon scattering cross sections are traditionally calculated~\cite{gandhi,lam349,lam372,lam383,kotikov}, is only consistent because of the remarkable cancellation between the quark and lepton contributions to triangle anomalies of gauged currents. This strongly suggests that in a more fundamental theory leptons and quarks should be interrelated~\cite{leptoquark}. Therefore, it seems not so unusual that at ultra-high energies the standard model may reflect some properties of this fundamental theory and the leptoquarks may manifest themselves in UHE neutrino interactions. 
Numerous theories lead to the possibility of quark--lepton interactions mediated by bosons called leptoquarks~\cite{lep1,lep2,lep3,lep4,lep5,lep6,lep7}. The analysis of this paper is based on the Buchm\"uller--R\"uckl--Wyler (BRW) leptoquark model~\cite{leptoquark}. 

Leptoquarks in different reactions including those initiated by neutrinos have been studied extensively in the literature~\cite{leptoquark,spira,blumlein1993, montalvo,vertex2,blumlein_main,lepto_nu1,zerwas04,lepto_nu2,zerwas94,mine,kosnik}.

Experimental searches for leptoquarks have been carried out in $e^+e^-$~\cite{exp1}, $ep$~\cite{exp2,hera}, $p\bar p$~\cite{exp3} and $pp$ collisions~\cite{exp4,exp5}. The most stringent limits to date on the first generation scalar and vector leptoquark masses are obtained by the H1 Collaboration at HERA in $ep$ collisions. For couplings corresponding to the leptoquark--quark--lepton vertex $g=0.3$, first generation leptoquarks with masses up to 800 GeV are excluded at 95\% confidence level~\cite{h1_13}. Limits on the scalar leptoquarks independent on $g$ are placed by the CMS Collaboration according to which masses below 640 GeV and 650 GeV are excluded, respectively for the first and second generation leptoquarks~\cite{exp4}. If the branching fraction of the leptoquark decay into a charged lepton and a quark is equal to unity, these limits increase to 830 GeV and 840 GeV, respectively.

\section{UHE neutrino--nucleon scattering in the Standard Model}
\label{model}

The (charged current, neutral current, total) cross section for deeply inelastic neutrino--nucleon scattering at neutrino energies $E_{\nu}=10^7-10^{12}$~GeV in the laboratory reference frame show a power rise with $E_{\nu}$ and is described by expressions of the following form~~\cite{gandhi,lam349,lam372,lam383}:

\begin{equation}
\sigma_{SM}(s)=Ns^{\lambda_0}\label{cross_1},
\end{equation}

where $s=2m_NE_{\nu}+m_N^2$ is the center-of-mass energy squared, $N$ and $\lambda_{0}$ are positive constants. Typically $\lambda_{0}\sim0.3$ and its particular value depends on the parton distribution functions (PDF) of the nucleon used in the evaluation of the cross section.

\begin{figure}
\centering
\resizebox{0.8\textwidth}{!}{%
\includegraphics{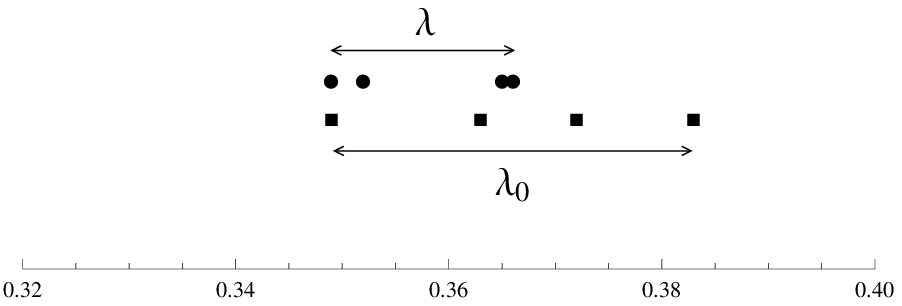}
}\caption{Comparison of the quantities $\lambda_0$ (squares) and $\lambda$ (circles) used in the parametrizations of the UHE neutrino--nucleon scattering cross sections and the quark densities, respectively. The values of $\lambda_0$ in ascending order are taken from~\cite{gandhi,lam349,lam372,lam383}. The values of $\lambda$ in ascending order are found  by using CTEQ6.6M~\cite{cteq66}, GJR08VFNS~\cite{gjr08}, GRV98-$\overline{\text{MS}}$~\cite{grv98}, GRV98-DIS~\cite{grv98} NLO PDFs at $Q^2=10^6$ GeV$^2$.}
\label{fig1}
\end{figure}

On the other hand, the quark densities in the nucleon at small Bjorken~$x$ relevant for the considered deep inelastic regime can be parametrized as~\cite{kotikov_lam}

\begin{equation}
xq(x)= Ax^{-\lambda}\label{pdf},
\end{equation}

where $A$ and $\lambda$ are some constants.  Next-to-leading (NLO) order fits of PDFs at $x=10^{-5}$ indicate that the parameter $\lambda$ is of the order of~0.3 for $Q^2>100$~GeV$^2$ and varies slowly with increasing $Q^2$~\cite{kotikov_lam}. It turns out that the ranges of variation of $\lambda_0$ and $\lambda$ substantially overlap. This fact is demonstrated in Fig.~\ref{fig1} where $\lambda_0$ from a series of papers is compared with $\lambda$ obtained by using different PDF sets at $Q^2=10^6$~GeV$^2$. The relation between $\lambda_0$ and $\lambda$ is a direct consequence of the structure of the standard deep inelastic neutrino scattering cross section:

\begin{equation}
\label{sm_cross} \sigma_{SM}(s)=
\frac{G_F^2}{\pi} \int_0^{1} dx\ \int_{0}^{xs} dQ^2\left(\frac{M_{W,Z}^2}{Q^2+M_{W,Z}^2}\right)^{2} \big[
q(x,Q^2) + {\bar q}(x,Q^2) (1-y)^2\big],
\end{equation} 

 where $y = Q^2/(x s)$. The effects of the $Q^2$-evolution of the quark distributions induce corrections of only about 20\%~\cite{insertion}. Then, performing the integration over $Q^2$ in~(\ref{sm_cross}) under the
approximation that one neglects the $Q^2$-evolution  yields~\cite{insertion}
\begin{equation}
\sigma_{SM}(s)\approx \frac{M_{W,Z}^2
G_F^2}{\pi} \int_{0}^1 d x
 \left[q(x)\left(\frac{{\hat s}}{1+{\hat s}}\right)  + {\bar q}(x)
\left(\frac{2}{{\hat s}}+1-2\left(\frac{1+{\hat s}}{{\hat s}^2}\right)
\ln(1+{\hat s})\right) \right],\label{sm_cross2}
\end{equation}

where ${\hat s}=xs/M_{W,Z}^2$ and the quark distribution functions are evaluated at some fixed scale. From~(\ref{sm_cross2}) one can see that the power-type behavior $xq(x)\propto x^{-\lambda}$ of the quark distributions
translates into a power-type rise with energy $\sigma_{SM}(s) \propto s^\lambda$.

\section{A phenomenological model of UHE neutrino--nucleon scattering}
\begin{figure}
\centering
\resizebox{1.\textwidth}{!}{%
\includegraphics{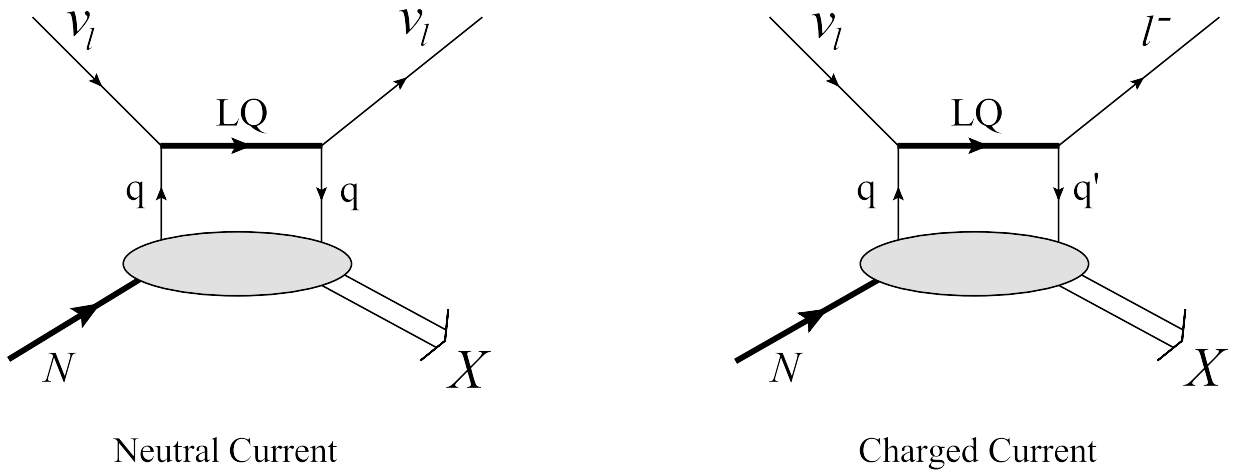}
}\caption{Neutrino--nucleon scattering through excitations of a leptoquark $LQ$. Left: the neutral current events. Right: the charged current events.}
\label{fig2}
\end{figure}

In addition to the standard picture of neutrino interactions, the necessity of the equivalence between $\lambda_0$ and $\lambda$ and the power law behavior of the cross section~(\ref{cross_1}) naturally stem  from a model within which UHE neutrino--nucleon scattering proceeds effectively through excitations of BRW-like leptoquarks coupling to neutrino--quark pairs as well as to charged lepton--quark pairs. This mechanism is schematically illustrated in Fig.~\ref{fig2}.   
The following subprocesses allowed by the BRW model can contribute to UHE $\nu N$ scattering:

\begin{equation}
\nu_l D \longrightarrow S^{(-1/3)}\put(2,5){\vector(4,1){25}}\put(2,2){\vector(4,-1){25}}\put(30,10){$\nu_l D$}\put(30,-10){$l^-U$,}\label{sub1}
\end{equation}

\begin{equation}
\nu_l \bar U \longrightarrow V^{(-2/3)}\put(2,5){\vector(4,1){25}}\put(2,2){\vector(4,-1){25}}\put(30,10){$\nu_l \bar U$}\put(30,-10){$l^-\bar D$.}\label{sub2}
\end{equation}  

Here $U=u,c$ (the top quark distribution in the nucleon is neglected)  and $D=d,s,b$; $S$ and $V$ stand for scalar and vector leptoquarks, respectively; the numbers in the brackets are the electric charges of the leptoquarks.






The cross section for each subprocess reads~\cite{leptoquark}:

\begin{equation}
\sigma^{(q)}_{LQ}(\hat s)=\frac{8\pi^2(2J+1)\Gamma(LQ\rightarrow\nu_l q)}{M_{LQ}}\delta(\hat s-M_{LQ}^2),\label{narrow}
\end{equation}

where $\hat s=xs$ is the total energy squared of the neutrino--quark collision, $M_{LQ}$ and $J$ are the mass and spin of the leptoquark, respectively, $\delta(z)$ is the Dirac delta function, $\Gamma(LQ\rightarrow\nu_l q)$ is the width of the decay $LQ\rightarrow\nu_l q$. The indices $q=d,s,b$ correspond to the subprocesses~(\ref{sub1}) and $q=\bar u,\bar c$ correspond to~(\ref{sub2}). Henceforth we take $\Gamma(S^{(-1/3)}\rightarrow\nu_l d)=\Gamma(S^{(-1/3)}\rightarrow\nu_l s)=\Gamma(S^{(-1/3)}\rightarrow\nu_l b)$ and  $\Gamma(V^{(-2/3)}\rightarrow\nu_l \bar u)=\Gamma(V^{(-2/3)}\rightarrow\nu_l \bar c)$.






In the considered deep inelastic regime it is fair to put $U(x)=\bar U(x)=D(x)=\bar D(x)$.  Then, convoluting~(\ref{narrow}) with the quark densities parametrized in the form of~(\ref{pdf}) yields

\begin{equation}
\sigma_{\nu N\rightarrow LQX}(s)=An_q\frac{8\pi^2(2J+1)\Gamma(LQ\rightarrow\nu_l q)}{M_{LQ}^{2\lambda+3}}s^{\lambda},\label{cross_sec2}
\end{equation}

where $n_q=3$ and $J=0$ if UHE neutrino--nucleon scattering proceeds through the subprocesses~(\ref{sub1}); $n_q=2$ and $J=1$ if UHE neutrino--nucleon scattering proceeds through the subprocesses~(\ref{sub2}). 

One can see from~(\ref{cross_sec2}) that our model reproduces the power rise of the cross section as $s^{\lambda}$ and thus explains the observation  $\lambda\approx\lambda_0$ displayed in Fig.~\ref{fig1}.

\subsection{The leptoquark mass}
The presented model  allows to evaluate the mass of the leptoquark in a way independent on its spin and electric charge.
Since  $\lambda$ depends on the momentum transfer squared $Q^2$~\cite{kotikov_lam} (in our case $Q^2\equiv\hat s$), the leptoquark mass will be given by the solution of the following equation:

\begin{equation}
\lambda(Q^2)=\lambda_0\label{equation}.
\end{equation}

From Fig.~\ref{fig1} one can already conclude that the mass is not far from 1~TeV.

\begin{figure}
\centering
\resizebox{1.0\textwidth}{!}{%
\includegraphics{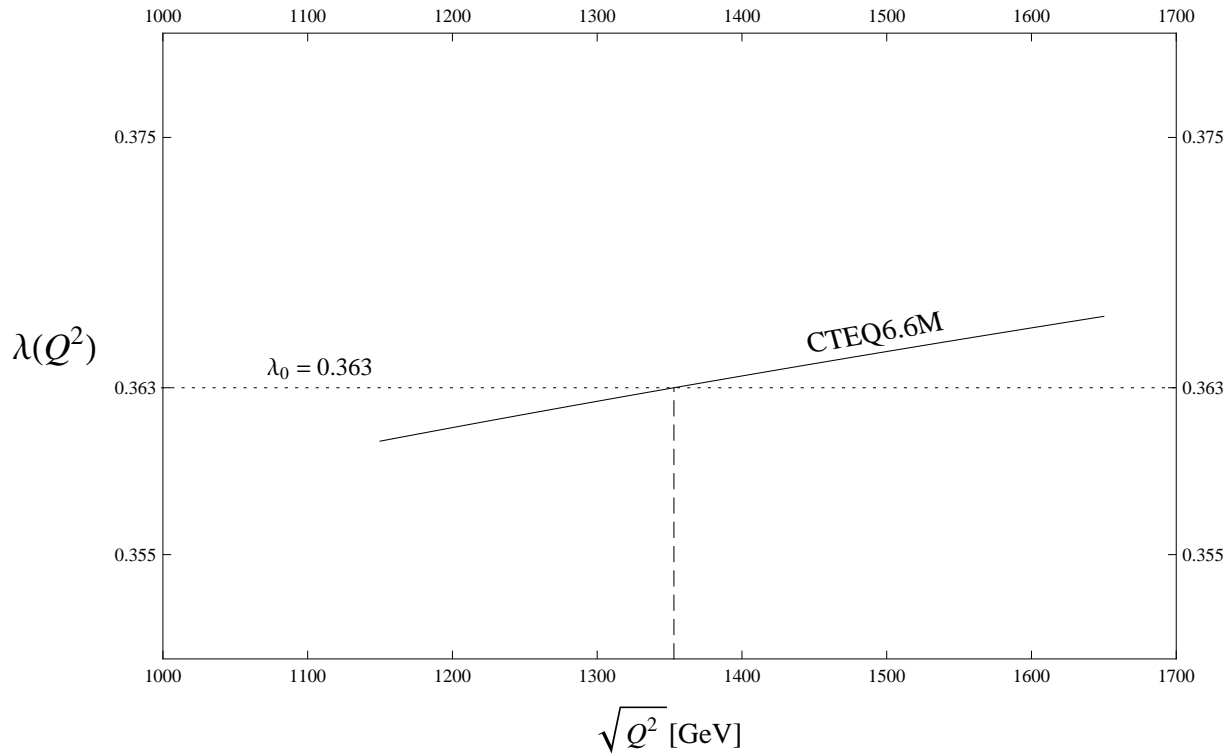}
}\caption{A graphical solution of the equation~(\ref{equation}). The solid line represents the function $\lambda(Q^2)$ found by adopting CTEQ6.6M PDFs. The value $\lambda_0=0.363$ is taken from~\cite{gandhi}. The solution corresponds to $\sqrt{Q^2}=M_{LQ}=1353$~GeV (vertical dashed line).}
\label{fig3}
\end{figure}

We have solved~(\ref{equation}) graphically at $\lambda_0=0.363$ taken from~\cite{gandhi} (see Fig.~\ref{fig3}). To find the function~$\lambda(Q^2)$ we adopted CTEQ6.6M parton distributions~\cite{cteq66}. This method gives within 17\% 

\begin{equation}
M_{LQ}=1353~\text{GeV}. 
\end{equation}

\subsection{The leptoquark decay widths}

Another consequence of the discussed model is the possibility to evaluate the leptoquark decay widths. 

By demanding

\begin{equation}
\sigma_{SM}(s)=\sigma_{\nu N\rightarrow LQX}(s)\label{demand},
\end{equation}

and using~(\ref{cross_1}), (\ref{cross_sec2}) and~(\ref{equation}) one obtains

\begin{equation}
\Gamma(LQ\rightarrow\nu_l q)=\frac{N M_{LQ}^{2\lambda+3}}{8\pi^2(2J+1)An_q}.\label{width}
\end{equation}

For instance, $\Gamma(V^{(-2/3)}\rightarrow\nu_l\bar U)=70.5\pm44.6$~GeV at $M_{LQ}=1353$ GeV, $N=7.84\times10^{-36}$~cm$^2$~\cite{gandhi}, $\lambda=0.363$~\cite{gandhi},  $A=0.223$ (the value of the parameter $A$ is found by using CTEQ6.6M~\cite{cteq66} and corresponds to $\lambda=0.363$). The leptoquark resonance turns out to be much wider than the massive gauge bosons of the Standard Model $W$ and $Z$. 
It is also obvious that 

\begin{equation}
\frac{\Gamma(LQ\rightarrow l^- q^{\prime})}{\Gamma(LQ\rightarrow\nu_l q)}=\frac{\sigma_{CC}(s)}{\sigma_{NC}(s)}\label{ratio}
\end{equation}

irrespective of the quantum numbers of the leptoquark through which UHE neutrino--nucleon scattering proceeds.
Here $\Gamma(LQ\rightarrow l^- q^{\prime})$ is the width of the channel $LQ\rightarrow l^- q^{\prime}$, $\sigma_{CC}(s)$ and $\sigma_{NC}(s)$ are the charged current and neutral current UHE neutrino--nucleon scattering cross sections. The Standard Model calculations show that the ratio~$\sigma_{CC}(s)/\sigma_{NC}(s)$ is approximately equal to 2.27 and very weakly depends on $s$ in the region $10^{7}~\text{GeV}\leq E_{\nu}\leq 10^{12}$ GeV~\cite{gandhi} so that $\Gamma(LQ\rightarrow l^- q^{\prime})/\Gamma(LQ\rightarrow\nu_l q)\approx2.27$. Meanwhile, the BRW model predicts $\Gamma(LQ\rightarrow l^- q^{\prime})=\Gamma(LQ\rightarrow\nu_l q)$. 

\section{Discovery potential of the LHC for leptoquarks}
The leptoquark mass $M_{LQ}=1353\pm230$ GeV predicted by the present analysis is within the mass range experimentally accessible at the CERN LHC  in leptoquark pair production~\cite{blumlein_main,zerwas04,aachen} as well as in single leptoquark production through generation of the equivalent lepton flux by the protons~\cite{zerwas94}. Let us consider the latter case in detail. 

\begin{figure}
\centering
\resizebox{0.8\textwidth}{!}{%
\includegraphics{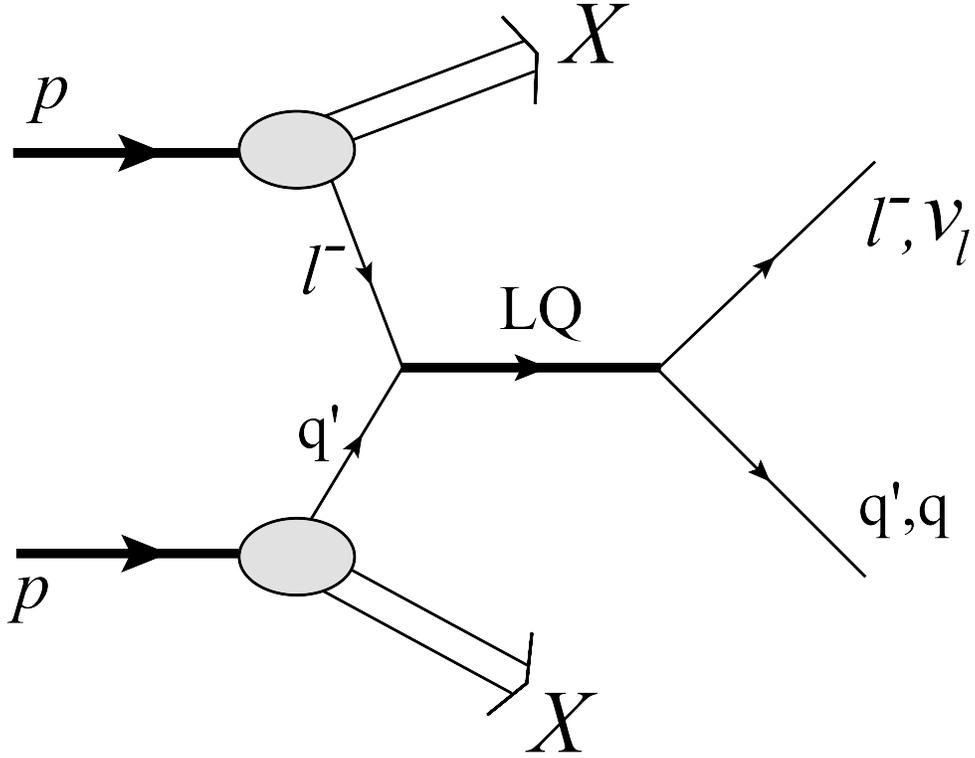}
}\caption{Schematic representation of the mechanism for producing single leptoquarks in
proton--proton collisions. The charged lepton appears from the equivalent lepton flux of the proton.}
\label{fig5}
\end{figure}

Leptoquarks can be produced singly at the LHC by splitting of photons emitted from the proton beam into lepton pairs. One of the leptons can then interact with a quark (antiquark) from the other proton beam and thereby produce a leptoquark. This mechanism is schematically illustrated in Fig.~\ref{fig5}. 

\begin{figure}
\centering
\resizebox{1.1\textwidth}{!}{%
\includegraphics{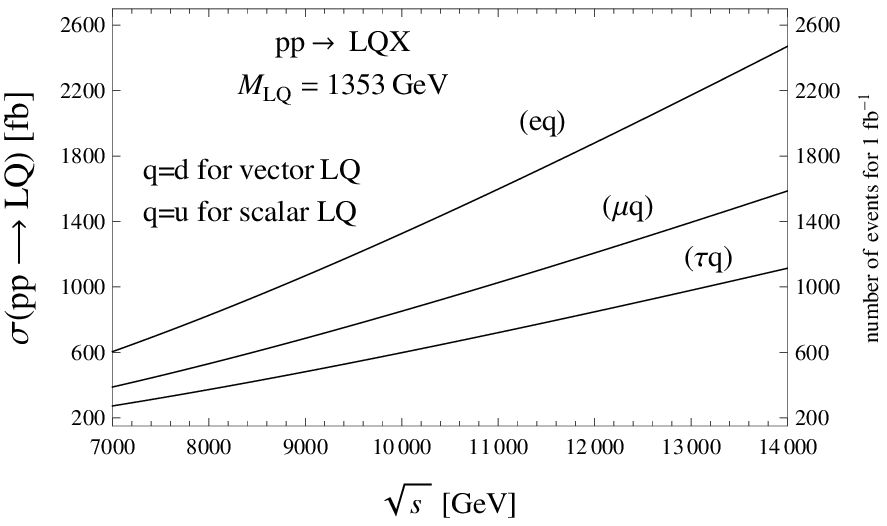}
}\caption{The cross sections for single inclusive production of the $(eu)$, $(\mu u)$, $(\tau u)$-type scalar leptoquarks and $(ed)$, $(\mu d)$, $(\tau d)$-type vector leptoquarks in $pp$ collisions as functions of the center-of-mass energy at $M_{LQ}=1353$ GeV. The number of the expected events for an integrated luminosity 1~$fb^{-1}$ is also shown.}
\label{fig8}
\end{figure}

Since the minimal value of $x$ probed in this process at the LHC is of the order of 0.01 ($x=Q^2/s=M^2_{LQ}/s$), the main contribution to the leptoquark production cross section will be from the valence $u$ or $d$ quark of the proton.  The BRW model allows two possibilities: 1) production of an ($ld$)-type vector leptoquark through the subprocess $l^+d\rightarrow V^{(+2/3)}$; 2) production of an ($lu$)-type scalar leptoquark through the subprocess $l^-u\rightarrow S^{(-1/3)}$.
Following~\cite{zerwas94},  we find the cross sections for single inclusive production of ($lq$)-type leptoquarks in $pp$ collisions valid for order of magnitude estimates:

\begin{equation}
\sigma(pp\rightarrow LQ)=\frac{\alpha^2|c|^2(2J+1)\Gamma(LQ\rightarrow l q)}{12M_{LQ}^3}\log{\left(\frac{M_{LQ}^2}{m_l^2}\right)}\log{\left(\frac{M_{LQ}^2}{m_q^2}\right)}\log^4{\left(\frac{s}{M^2_{LQ}}\right)},\label{cross_lq}
\end{equation}

where $\alpha$ is the fine structure constant, $m_l$ and $m_q$ are the masses of the charged lepton and quark, respectively. The parameter $|c|=0.16$ appears in estimating the proton structure function $F_2$ (see~\cite{zerwas94}). The values of the cross sections for both leptoquarks $S^{(-1/3)}$ and $V^{(+2/3)}$ are close each to other  because in the proton $u(x)\approx2d(x)$, so that for order of magnitude estimates one can take $\sigma(pp\rightarrow S^{(-1/3)})=\sigma(pp\rightarrow V^{(+2/3)})$.
Figure~\ref{fig8} displays the energy dependence of the cross sections for the leptoquark mass $M_{LQ}=1353$ GeV.  

Single leptoquark production may be relatively easy to analyze experimentally due to a leptoquark decays into a charged lepton plus a quark. In this case, the final state has simple topology  formed by an identified charged lepton plus jet, both at large transverse momenta ($p_T\sim M_{LQ}/2$). The main contribution to the background arises from $W(\rightarrow l\nu_l)$ plus jet final states. Unlike  leptoquark decays, the transverse momenta of leptons and jets are not balanced in the background events and the latter can be well separated out. The neutrino decay channels of the leptoquarks can also be exploited but the rejection of the corresponding background $Z(\rightarrow\nu\bar\nu)$+jet  is a more difficult task since one has to reconstruct the invariant mass from missing momentum and energy~\cite{zerwas94}.

According to the predicted leptoquark decay widths and the cross sections, a singly produced leptoquark will manifest itself at the LHC as a relatively wide resonance with $\Gamma\sim100$ GeV residing in the mass range between 1123~GeV and 1583~GeV. For the energy $\sqrt{s}=7$ TeV and an integrated luminosity of 4.7 fb$^{-1}$ already reached at the LHC, the expected number of leptoquark events with the topologies ($e+\text{jet}$) and ($\mu+\text{jet}$) is $\sim10^3$  and grows to values of the order of $10^4$ for $\sqrt{s}=14$ TeV at the same luminosity.

In the case of scalar and vector leptoquark pair production through gluon--gluon fusion and quark--antiquark annihilation at the LHC, the number of expected events is much less ($\sim10^2$) for $\sqrt{s}=14$~TeV at the luminosity 10~fb$^{-1}$~\cite{blumlein_main} (here the minimal vector coupling between the vector leptoquarks and gluons is assumed).

\section{Conclusions}
Relying on the behavior of the energy dependence of UHE neutrino--nucleon scattering cross sections and parton distribution functions at small values of the Bjorken variable~$x$, we have  proposed a model within which UHE neutrino--nucleon scattering effectively proceeds through excitations of BRW-like leptoquarks in neutrino--quark subprocesses. This model allows to estimate masses as well as the decay widths of the involved leptoquarks. Our approach predicts the leptoquark mass $1353\pm230$ GeV in a way independent on the leptoquark quantum numbers. It is also found that the decay widths of the leptoquarks are of the order of 100 GeV. The discovery potential of the LHC for the leptoquarks is studied in detail. In particular, the cross sections for single production of ($ld$)-type vector leptoquarks through the subprocess $l^+d\rightarrow LQ^{(+2/3)}$ and ($lu$)-type scalar leptoquarks through the subprocess $l^-u\rightarrow LQ^{(-1/3)}$ in $pp$ collisions at the LHC energies are evaluated. According to our predictions, a singly produced leptoquark will manifest itself at the LHC as a relatively wide resonance with $\Gamma\sim100$ GeV residing in the mass range between 1123~GeV and 1583~GeV. For the energy $\sqrt{s}=7$ TeV and an integrated luminosity of 4.7 fb$^{-1}$ already reached at the LHC, the expected number of leptoquark events with the topologies ($e+\text{jet}$) and ($\mu+\text{jet}$) is $\sim10^3$  and grows to values of the order of $10^4$ for $\sqrt{s}=14$ TeV at the same luminosity. 

It is also interesting to discuss the manifestation of the leptoquarks in UHE neutrino scattering processes. The BRW model gives $\Gamma(LQ\rightarrow l^- q^{\prime})=\Gamma(LQ\rightarrow\nu_l q)$. The realization of this case within our model will be experimentally observed as equivalence between $\sigma_{CC}(s)$ and $\sigma_{NC}(s)$, while the Standard Model predicts  $\sigma_{CC}(s)/\sigma_{NC}(s)\approx2.27$. 
The BRW model accommodates leptoquarks which couple only to neutrino--quark pairs so that $\Gamma(LQ\rightarrow l^- q^{\prime})=0$. Such quarks, if responsible for UHE neutrino scattering, will lead to "switching off" the charged current channels at $E_{\nu}\rightarrow\infty$. This may have important consequences for physics of UHE cosmic ray neutrinos.

\vskip 0.5cm

\acknowledgements

This work was supported in part by the Russian Foundation for Basic Research (grant 11-02-12043), by the Program for Basic Research of the Presidium of the Russian Academy of Sciences "Fundamental Properties of Matter and Astrophysics" and by the Federal Target Program  of the Ministry of Education and Science of Russian Federation "Research and Development in Top Priority Spheres of Russian Scientific and Technological Complex for 2007-2013" (contract No. 16.518.11.7072).


\newpage

\newpage

\begin{thebibliography}{00}
\bibitem{gandhi}
R. Gandhi, C. Quigg, M. H. Reno, I. Sarcevic,  	Phys. Rev. D {\bf58}, 093009 (1998).

\bibitem{lam349}
R. Basu, D. Choudhury, S. Majhi, JHEP {\bf10}, 012 (2002).

\bibitem{lam372}
M. Gl\"uck, S. Kretzer, E. Reya, Astropart. Phys. {\bf11}, 327 (1999).

\bibitem{lam383}
D. K. Choudhury, P. K. Dhar,  arXiv:1103.3788. 

\bibitem{kotikov}
R. Fiore {\it et al.}, Phys. Rev. D {\bf71}, 033002 (2005).



\bibitem{leptoquark}
W. Buchm\"uller, R. R\"uckl, D. Wyler, Phys. Lett. B {\bf191}, 442 (1987); Erratum ibid. B {\bf448}, 320 (1999). 

\bibitem{lep1}
H. Georgi, S. Glashow, Phys. Rev. Lett. {\bf32}, 438
(1974).

\bibitem{lep2}
J. Pati, A. Salam, Phys. Rev. D {\bf10}, 275 (1974).

\bibitem{lep3}
G. Senjanovic, A.Sokorac, Z. Phys. C {\bf20}, 255 (1983).

\bibitem{lep4}
B. Schrempp, F. Schrempp, Phys. Lett. B {\bf153}, 101 (1985).

\bibitem{lep5}
E. Witten, Nucl. Phys. B {\bf258}, 75 (1985).

\bibitem{lep6}
P. Frampton, B.-H. Lee, Phys. Rev. Lett. {\bf64}, 619
(1990).

\bibitem{lep7}
P. Frampton, T. Kephart, Phys. Rev. D {\bf42}, 3892
(1990). 


\bibitem{spira}
A. Djouadi, T. K\"ohler, M. Spira, J. Tutas, Z. Phys. C {\bf46}, 679 (1990).

\bibitem{blumlein1993}
J. Bl\"umlein, R. R\"uckl, Phys. Lett. B {\bf304}, 337 (1993).

\bibitem{montalvo}
J. E. Cieza Montalvo, O. J. P. Eboli, Phys. Rev. D {\bf47}, 837 (1993).

\bibitem{vertex2}
T. M. Aliev, E. Iltan, N. K. Pak, Phys. Rev. D {\bf54}, 4263 (1996).

\bibitem{blumlein_main}
J. Bl\"umlein, E. Boos, A. Kryukov, Z. Phys. C {\bf76}, 137 (1997).

\bibitem{lepto_nu1}
M. A. Doncheski, R. W. Robinett, Phys. Rev. D {\bf56}, 7412 (1997).

\bibitem{zerwas04}
M. Kr\"amer, T. Plehn, M. Spira, P.M. Zerwas,  	Phys. Rev. D {\bf71}, 057503 (2005).

\bibitem{lepto_nu2}
L. A. Anchordoqui, C. A. G. Canal, H. Goldberg, D. G. Dumm, F. Halzen, Phys. Rev. D {\bf74}, 125021  (2006).

\bibitem{zerwas94}
J. Ohnemus, S. Rudaz, T.F. Walsh, P.M. Zerwas, Phys. Lett. B {\bf334}, 203 (1994).

\bibitem{mine}
I. Alikhanov, Phys. Lett. B {\bf717}, 425 (2012).

\bibitem{kosnik}
I. Dorsner, J. Drobnak, S. Fajfer, J. F. Kamenik, N. Kosnik,
JHEP {\bf11}, 002 (2011).



\bibitem{exp1}
OPAL Collaboration, Eur. Phys. J. C {\bf31}, 281 (2003).

\bibitem{exp2}
H1 Collaboration, Phys. Lett. B {\bf704}, 388 (2011).

\bibitem{hera}
ZEUS Collaboration, arXiv:1205.5179.

\bibitem{exp3}
D0 Collaboration, Phys. Rev. D {\bf84}, 071104(R) (2011).

\bibitem{exp4}
CMS Collaboration, Phys. Rev. D {\bf86}, 052013 (2012).

\bibitem{exp5}
ATLAS Collaboration, Phys. Lett. B {\bf709}, 158 (2012).

\bibitem{h1_13}
D. M. South, arXiv:1302.3378.


\bibitem{kotikov_lam}
A. V. Kotikov, arXiv:1212.3733.

\bibitem{insertion}
D. A. Dicus, S. Kretzer, W. W. Repko, C. Schmidt,
Phys. Lett. B {\bf514}, 103 (2001).

\bibitem{cteq66}
P. M. Nadolsky et al., Phys. Rev. D {\bf78}, 013004 (2008).

\bibitem{gjr08}
M. Gl\"uck, P. Jimenez-Delgado, E. Reya, C. Schuck, Phys. Lett. B {\bf664}, 133 (2008).

\bibitem{grv98}
M. Gl\"uck, E. Reya, A. Vogt,  	Eur. Phys. J. C {\bf5}, 461 (1998).

\bibitem{aachen}
N. Harnew, Proceedings “Large Hadron Collider”, Aachen 1990, CERN 90–10.

\end{thebibliography}
\end{document}